\documentclass[preprint]{emulateapj} 
\usepackage{todonotes}
\usepackage{csvsimple}
\usepackage{longtable}
\usepackage{amsmath}
\usepackage{booktabs}
\usepackage{ gensymb }
\usepackage{verbatim}

\newcommand{\kms}{km~s$^{-1}$}
\newcommand{\ms}{m~s$^{-1}$}
\newcommand{\Hipparcos}{{\it Hipparcos}}

\begin{document}

\title{Extracting Radial Velocities of A- and B-type Stars \\ from Echelle Spectrograph Calibration Spectra}  
\shorttitle{Radial Velocities of A- and B-type Stars}
\author{Juliette C. Becker\altaffilmark{1,2,3}, John Asher Johnson\altaffilmark{4,5}, Andrew Vanderburg\altaffilmark{3,4}, Timothy D. Morton\altaffilmark{6}} 
\shortauthors{Becker et al. }
\altaffiltext{1}{Department of Astronomy, University of Michigan, 1085 S University Ave, Ann Arbor, MI 48109}
\altaffiltext{2}{Cahill Center for Astronomy and Astrophysics, California Institute of Technology, 1200 E. California Blvd., Pasadena, CA 91125}
\altaffiltext{3}{NSF Graduate Research Fellow}
\altaffiltext{4}{Harvard-Smithsonian Center for Astrophysics, 60 Garden St., Cambridge, MA 02138}
\altaffiltext{5}{David \& Lucile Packard Fellow}
\altaffiltext{6}{Department of Astrophysical Sciences, 4 Ivy Lane, Peyton Hall, Princeton University, Princeton, NJ 08544}

\email{jcbecker@umich.edu}

 \begin{abstract} 
We present a technique to extract radial velocity measurements from echelle spectrograph observations of rapidly rotating stars ($V\sin{i} \gtrsim 50$~\kms). This type of measurement is difficult because the line widths of such stars are often comparable to the width of a single echelle order. To compensate for the scarcity of lines and Doppler information content, we have developed a process that forward--models the observations, fitting the radial velocity shift of the star for all echelle orders simultaneously with the echelle blaze function. We use our technique to extract radial velocity measurements from a sample of rapidly rotating A-- and B--type stars used as calibrator stars observed by the California Planet Survey observations. We measure absolute radial velocities with a precision ranging from 0.5--2.0~\kms\ per epoch for more than 100 A- and B-type stars. In our sample of 10 well-sampled stars with radial velocity scatter in excess of their measurement uncertainties, three of these are single--lined binaries with long observational baselines. From this subsample, we present detections of two previously unknown spectroscopic binaries and one known astrometric system. Our technique will be useful in measuring or placing upper limits on the masses of sub-stellar companions discovered by wide--field transit surveys, and conducting future spectroscopic binarity surveys   and Galactic space--motion studies of massive and/or young, rapidly--rotating stars.

\end{abstract} 


  \maketitle

\section{Introduction}

Stellar radial velocity (RV) measurements have become increasingly precise over the past 30 years due to the advent and development of high--resolution spectrographs equipped with digital detectors \citep{1981SPIE..290..215C}, including HIRES at Keck \citep{1994SPIE.2198..362V, 2010ApJ...721.1467H}; and particularly with the construction of environmentally-stabilized spectrometers such as the HARPS-South and -North spectrographs \citep{2003Msngr.114...20M, 2012SPIE.8446E..1VC}, SOPHIE at Haute-Provence \citep{2009A&A...505..853B}, CHIRON at CTIO \citep{2010SPIE.7735E..4GS}, and the Planet Finder Spectrograph (PFS) at Magellan \citep{2006SPIE.6269E..31C, 2010SPIE.7735E..53C}. 
While the discovery and characterization of exoplanets has been the driving scientific motivation behind these developments \citep[e.g.][]{1995Natur.378..355M, 1999ApJ...526..916B, 2004ApJ...617..580B, 2012Natur.491..207D}, increased measurement precision has also led to significant advances in understanding stellar binarity, particularly around Sun-like stars \citep{1991A&A...248..485D,1992ApJ...396..178F,2010ApJS..190....1R}.

However, the stability of a given spectrometer is only part of what enables high radial velocity precision. The attainable Doppler precision also depends greatly on the type of star observed. Measurements at the highest attainable precision today, levels at or below 1~\ms, can only be performed on stars with spectra that contain many sharp spectral lines.  As a result, most RV--based planet surveys have been restricted to F-,G-,K-,and M-type dwarf stars, which rotate slowly and display numerous fine spectral features.  

\begin{figure}[htbp] %  figure placement: here, top, bottom, or page
   \centering
   \includegraphics[width=3.4in]{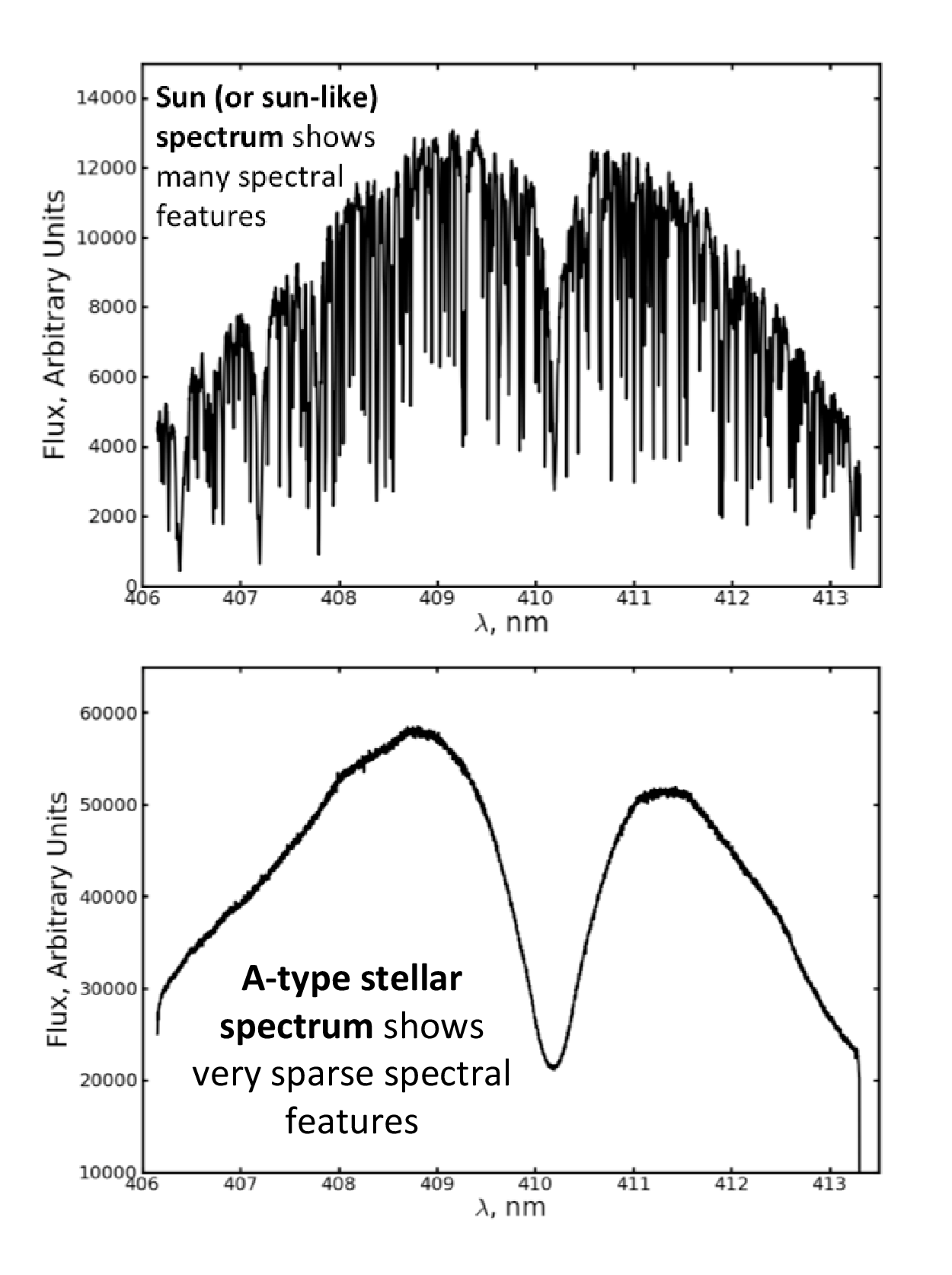} 
   \caption{Radial velocity measurements using A- and B-type stellar spectra are hindered by rotational broadening of their observed spectral features. \textcolor{black}{Here, we see this effect as illustrated by the $H\delta$ 410.1 nm Balmer line.}{\em Upper panel}: One HIRES echelle order of the Solar spectrum measured by observing reflected Sunlight from the asteroid Vesta. This spectrum is representative of those of low-mass stars observed by the CPS program. {\em Lower panel}: the same HIRES order, this time showing an observation of an A--type star, HR\,6827. This rapidly-rotating star has hundreds of times fewer spectral features than are seen in the Solar spectrum. The high-mass star cannot be analyzed in the same way as a Sun-like due to its broad spectral features, which are significant fraction of their echelle orders.}
   \label{comparison}
\end{figure}

On the other hand, more massive A- and B-type stars have hotter atmospheres and exhibit fewer absorption features. Also, because these hot stars lack convective outer layers, they retain most of their primordial angular momentum, and what few spectral features they show are highly rotationally broadened. For these reasons, rapidly--rotating hot and massive stars have nearly featureless blackbody spectra, showing only very broad hydrogen and helium absorption lines, as illustrated in Figure \ref{comparison}. Rotational smearing also affects young stars of all masses if they have not yet lived long enough to have experienced sufficient magnetic braking. It is thus much more challenging to obtain precise RVs for rapidly rotating stars from high--resolution echelle observations.

At the same time, their nearly featureless spectra make hot stars excellent calibrators for measuring and removing telluric absorption features, and as  calibrators for Doppler surveys \citep[as well as for instrumental tests, as in][]{2013PASP..125..511S}. These ``blackbodies in the sky'' are excellent calibrators of the transmission functions of absorption cells used as wavelength references, and as means of measuring the spectrometer's instrumental profile for surveys using gas absorption cells. As a result, there exists a large library of high--resolution spectra of hot stars obtained as calibrators of high-precision, gas-cell calibrated Doppler surveys such as the California Planet Survey (CPS). 

While this library was obtained for calibration purposes rather than as a scientific data product, it serendipitously provides the opportunity to conduct a radial velocity survey of hot stars. Multiplicity studies of high mass stars are important to constrain models of their formation \citep{2005MNRAS.362..915B, 2007ARA&A..45..481Z}. Some notable massive-star radial velocity studies include those of \cite{2005A&A...443..337G}, who studied the multiplicity of A- and F-type dwarfs with rapid rotation rates, \cite{2012MNRAS.424.1925C}, who examined the binary fraction among B- and O-type stars, and \cite{2010ApJ...722..605H}, who examined radial velocities as a larger-scale effort to measure the projected rotational velocities of massive stars. 

In addition to binarity surveys, absolute radial velocities (velocities measured with respect to the solar system barycenter) offer the ability to study the local motion and bulk flow of stars in the Galaxy, informing cluster dynamics, providing formation insights to the formation histories of visible stars, and providing information about the assembly of the Milky Way and studies of open clusters \citep[e.g.][]{2010AN....331..474F, 2009A&A...498..949M}.
Absolute radial velocities are also needed to calibrate other measurements. For example, the {\em Hipparcos}-Gaia Hundred-Thousand Proper-Motion survey, which aims to find the proper motions for over $\sim10^5$ stars over a 23-year baseline, requires radial velocity measurements of its target stars to account for acceleration that might be affecting the proper motion measurements \citep{2012A&A...546A..61D}. 

While echelle spectra of rapidly rotating A- and B-type stars show very few absorption features, the high signal-to-noise (SNR) and the highly oversampled nature of the spectral features of their spectra should, in principle, provide radial velocity better than 1 \kms (see Appendix \ref{precision}). This precision allows both absolute measurements to measure the space motions of these bright stars, as well as relative radial velocity measurements to search for binary companions. 

In Section \ref{methody} we present a new analysis technique to extract RV measurements from echelle spectra of rapidly rotating stars, for both differential (Section~\ref{fitting_rv})  and absolute (Section \ref{absrvs}) velocities. Section \ref{highscatter} presents the results of applying our method to a large number of archival high-resolution spectra of A- and B-type stars. These observations were obtained for use as calibrators by the California Planet Survey RV planet search program at Keck Observatory, over a span of 8 years since the HIRES detector upgrade. We achieve a typical precision of 1 \kms, and recover the orbital motion of several known astrometric or spectroscopic binaries. Additionally, we detect long-term radial velocity trends for two stars (HR\,5867 and HR\,8028) and spectroscopically confirm the astrometric binary HR\,3067.

\section{Observations and Analysis}
\label{methody}

\subsection{Data Collection}

The data presented herein were collected with the HIgh Resolution Echelle Spectrograph (HIRES) on the Keck~I telescope \citep{1994SPIE.2198..362V}. HIRES was operated in the standard CPS observing mode with the red-optimized grating with a spectral resolving power of $\Delta \lambda/\lambda \approx 55,000$. Across the 8 years of observations, various slit masks, or ``deckers'' were used , including C1, C5, B1 and B5 \footnote{{\url http://www2.keck.hawaii.edu/inst/hires/manual2.pdf}}. Because the targets are so bright and since the rotational broadening of the stars is large enough that all features are resolved, the different observing modes have little effect on the final results.

HIRES has three charge coupled devices (CCDs), each of which covers a different wavelength range of the spectrum. Colloquially, these CCDs are referred to as the blue (364.3--479.5~nm), green (497.7--642.1~nm), and red (654.3--799~nm) chips. These CCDs have twenty-three, sixteen, and ten orders, respectively, with each constituent order containing 4020 pixels. We reduce the HIRES CCD images using the standard CPS method of using an optimal extraction technique to trace spectral orders on the two-dimensional echelle image, rectifying the orders, and then summing pixels in columns to obtain a one--dimensional spectrum for each order.

The observations of rapidly rotating calibrators were often, but not always, made with the iodine cell in the light path in order to measure the instrumental profile (line-spread function) from the sharp iodine absorption features \citep{1996PASP..108..500B, 2006ApJ...647..600J}. In our analysis, we take advantage of the simultaneous iodine reference to determine the spectrograph's wavelength solution. In the cases of exposures taken without the iodine cell, we search for the nearest observation in time taken with the iodine cell and use its wavelength solution instead.  We note, however, that the region containing iodine lines only spans about 100~nm about on the green chip between roughly 500~nm and 600~nm, and the majority of the spectral features characteristic of hot, rapidly rotating stars (in particular, hydrogen and helium lines) are located on the blue chip. Using the iodine--derived wavelength solution requires extrapolation to the rest of the HIRES bandpass, which we describe in Section \ref{wvls}.

\subsection{Wavelength Solution}
\label{wvls}

HIRES is not an environmentally--stabilized instrument like other precise RV instruments, so the spectrograph's wavelength solution drifts over the course of a night at the level of 1--2~\kms (roughly a pixel). The CPS program circumvents this problem by passing starlight through an iodine gas cell, which imprints a wavelength reference spectrum onto the intrinsic stellar spectrum \citep{1995PASP..107..966V}. The simultaneous iodine reference permits wavelength solutions to a precision better than a fraction of a m~s$^{-1}$, significantly more precise than is necessary for RV measurements of rapidly rotating hot stars. Unfortunately, this wavelength solution is only measurable between roughly 500~nm and 600~nm, where there is significant iodine absorption and a lack of strong telluric absorption features.

We found that even though the iodine wavelength solution was only calculated over a small region, it was possible to extrapolate the wavelength solution to other spectral regions with a precision of better than 400~\ms. This is possible because the orientation of the three CCDs is such that all orders (on all three chips) are parallel, with the response functions by pixel remaining consistent (to 0.4 \kms)  between orders. The distance between chips is only 6-7 pixels\footnote{\url{ http://www2.keck.hawaii.edu/inst/hires/hires\_data.pdf}}, so the wavelength mapping for one CCD is closely matched by the neighboring CCDs.  

\begin{figure}[htbp] %  figure placement: here, top, bottom, or page
   \centering
   \includegraphics[width=3.4in]{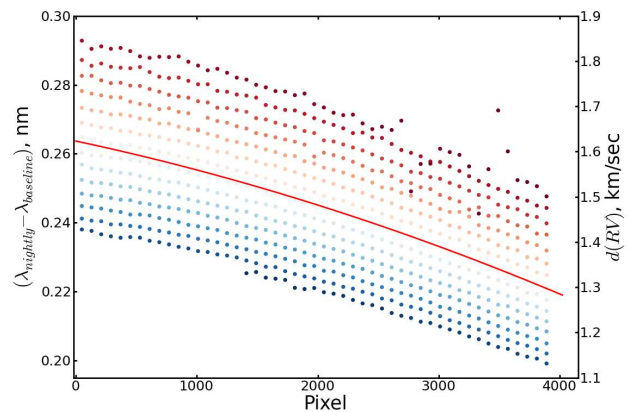} 
   \caption{Difference between HIRES wavelength solution, \textcolor{black}{$\lambda_{\rm nightly}$, from one epoch compared to the} nominal \textcolor{black}{wavelength solution
   %, $\lambda_{\rm baseline}$ 
   from an arbitrary reference epoch} as a function of pixel and order number for the green chip. Each order is represented by a different color, from the bluest to reddest order. The dots are measurements of $\delta \lambda$, the deviation in wavelength from the nominal solution, and the solid line is the best-fitting solution \textcolor{black}{as described by Equation \ref{extraps} for an arbitrary order $n$}. The offset from zero indicates that the spectrograph's wavelength zero--point has shifted from the nominal value \textcolor{black}{by about 1.4 \kms\ at this epoch, and the slope---wavelength zeropoint as a function of pixel---is described by coefficients $B$ and $C$ in Equation \ref{extraps}.}}
   \label{WLS}
\end{figure}

We fit the wavelength solution from the iodine region on the green chip with the following model: 

\begin{equation}
\lambda_{extrap}(i, n) = A + B \times i + C \times i^2 + D \times n 
\label{extraps}
\end{equation}

\noindent where $n$ is the order number, $i$ is the pixel number in the dispersion direction, and $A$, $B$, $C$, and $D$ are the fitted coefficients. This model does a good job of describing the dependence in the dispersion direction, but the simple linear dependence of wavelength on order number is only adequate to describe the wavelength solution to a precision of $\simeq$ 100 \ms\ on the green chip. Fitting only a linear dependence on order number, however, allows us to extrapolate the wavelength solution without the problem of a higher order polynomial fit diverging quickly. \textcolor{black}{For an example wavelength solution, shown in Figure \ref{WLS}, the nominal solution must be corrected by values varying by 400 \ms\ between the bluest and reddest orders. The exact value of this deviation varies by observation, but the values given here are typical values. The correction computed from the green chip can then be applied to the blue chip, resulting in a wavelength solution with at least 400 \ms\ better precision between orders on the far edges of the chip and 1.4 \kms\ better precision compared to the nominal wavelength solution for the entire run.}

We show the result of one of the fits in Figure \ref{WLS} and compare the fit to the data in the iodine region. The uncorrected dependence on order number is evident, but the typical errors introduced are small. Due to the large drifts in the wavelength solution of the spectrograph, the extrapolated solution is of higher quality in general than the nightly solution, ($\lambda_{\rm nightly}$), which is calculated from a Thorium-Argon lamp exposure at the beginning of the night. We therefore use the extrapolated wavelength solutions in our analysis hereafter.

\subsection{Continuum Shape}
\label{contshape}

Unlike traditional Doppler techniques developed for F-, G-, K-, and M-stars, which perform analysis on continuum normalized spectra, the peculiarities of our hot, rapidly rotating stellar sample require us to simultaneously fit for radial velocities with the spectrograph's blaze function. Hot, rapidly rotating stars like those considered in this study have very broad spectral features, some of which have line widths that are a significant fraction($\sim$ 10-20\%) of the width of a HIRES echelle order (typically 5~nm in the blue). Ignoring this would introduce biases caused by the degeneracy between the overall flux level and the location of spectral lines, \textcolor{black}{preventing effective normalization to the continuum.}
            
The shape of the continuum for each spectral order in echelle spectrographs is a blaze function determined by the spacing of grooves on the diffraction grating.  The Fourier transform of the shape of each groove on the grating results in a sinc ($\sin(x)/x$) function, the first maximum of which is known as the blaze function. The shapes of the spectral orders are similar due to their common physical origin, but other spectrograph optical effects result result in small changes between orders.

We take advantage of the fact that the echelle orders have similar blaze function shapes when modeling them in our radial velocity fits, and model the continuum shape of the continuum as a function of both order number and pixel number in the dispersion direction. 
Essentially, the continuum level of each spectral order is a slice from one continuous, three--dimensional function $F(i, n)$, where $F$ is the flux level, $n$ is to the order number, and $i$ is to the pixel number in the dispersion direction. 

We experimented with various functional forms for $F(i,n)$ by fitting to flat field exposures -- that is, calibration exposures taken when the HIRES slit was illuminated with a quartz lamp continuum source. We settled on the following form for $F(i,n)$: 

\begin{equation}
F(i,n) = c_{0} i^{2} + c_{1} i + c_{2} + c_{3} n + c_{4} n^{2} + c_{5} i n,
\label{contin_level}
\end{equation}

\noindent where the coefficients, $\{c_j\}$, are free parameters.

\subsection{Fitting Procedure}
\label{fitting_rv}

We measure radial velocities by simultaneously fitting a model to the continuum shape and Doppler shift of each spectrum. We start by selecting the first  observation for a given star and set this spectrum to be our stellar template spectrum, analogous to the deconvolved intrinsic stellar spectrum used in iodine cell Doppler analysis \citep{1996PASP..108..500B}. However, since the instrumental line-broadening is negligible to the rotational broadening, no deconvolution is required. 

We then use a Levenberg--Marquardt \citep{2002nrc..book.....P} least squares technique to find the best--fitting Doppler shift for each observation of a particular star. The Levenberg-Marquardt algorithm is relatively robust, but can sometimes get stuck in local extrema in the function it is minimizing or maximizing, so we take care to find good initial guesses for the fit parameters. We first estimate the Doppler shift by performing a cross correlation between the template spectrum with the observation on one particular spectral order, the one containing the H-$\gamma$ line at 434.047~nm. We estimate the shape of the continuum using our fits to the flat field lamp as described in Section \ref{contshape}. Once we have initial guesses, we perform the Levenberg-Marquardt maximization on the following log--likelihood (where $\ln{L} \propto -\chi^2$) function: 

\begin{equation}
\ln{L} = \sum_{i=0}^{\textcolor{black}{N-1}} \left[ -\frac{1}{2}\ln{(2 \pi \sigma_i )} - \frac{1}{2} \left(\frac{\mathcal{I}_{i} - \mathcal{I}_{m}}{\sigma_{i}} \right)^{2}\right]
\label{lhood}
\end{equation}

\noindent where $\sigma_i$ is the error in flux on each pixel, \textcolor{black}{$N$ is the total number of pixels,} and $\mathcal{I}_{i}$ is the flux of the datum spectrum at pixel $i$. The model $\mathcal{I}_m$ is given by the following expression: 

\begin{equation}
\begin{split}
\mathcal{I}_{m}(i, n) = \rm{F}_{\textcolor{black}{\mathbf{ratio}}}(i,n) &\cdot S_m \left[\lambda_{\rm{extrap}}(i, n)\left(1+\textcolor{black}{\frac{V_{\rm Dop}}{c}}\right) \right] 
\end{split}
\label{model}
\end{equation}

\noindent where $i$ is pixel number, $n$ is order number, $\rm{F}_{\textcolor{black}{ratio}}(i,n)$ is the ratio between two continuum levels which share the functional form of Equation \ref{contin_level}, and $S_m[i, n]$ is the original flux level of the first observed spectrum, which is used as the model.
It is not necessary to include the convolution kernel in the fit for relative radial velocities, as the rotation rate and instrument profile are expected to be constant between successive observations. If desired, the broadening can be fit by convolving Equation \ref{model} with the kernel as described in Equation \ref{kkernel}.

After the likelihood maximization, we extract the best--fitting Doppler shift parameter ($V_{\rm Dop}$). We treat the other model inputs as nuisance parameters. Finally, we apply a barycentric correction to the best-fitting Doppler shift to correct for the Earth's motion with respect to the target star. The barycentric correction depends primarily on the declination of the target, reaching a maximum for targets on the ecliptic. Higher--order contributions to the barycentric correction can be safely ignored at our target precision of $\sim1$~\kms. We computed the barycentric correction with a python adaptation of the \texttt{baryvel} code \citep{1980A&AS...41....1S} with errors much smaller than our expected precision.

\subsection{Absolute Radial Velocities}
\label{absrvs}

\begin{figure*}[htbp] %  figure placement: here, top, bottom, or page
   \centering
   \includegraphics[width=7.2in]{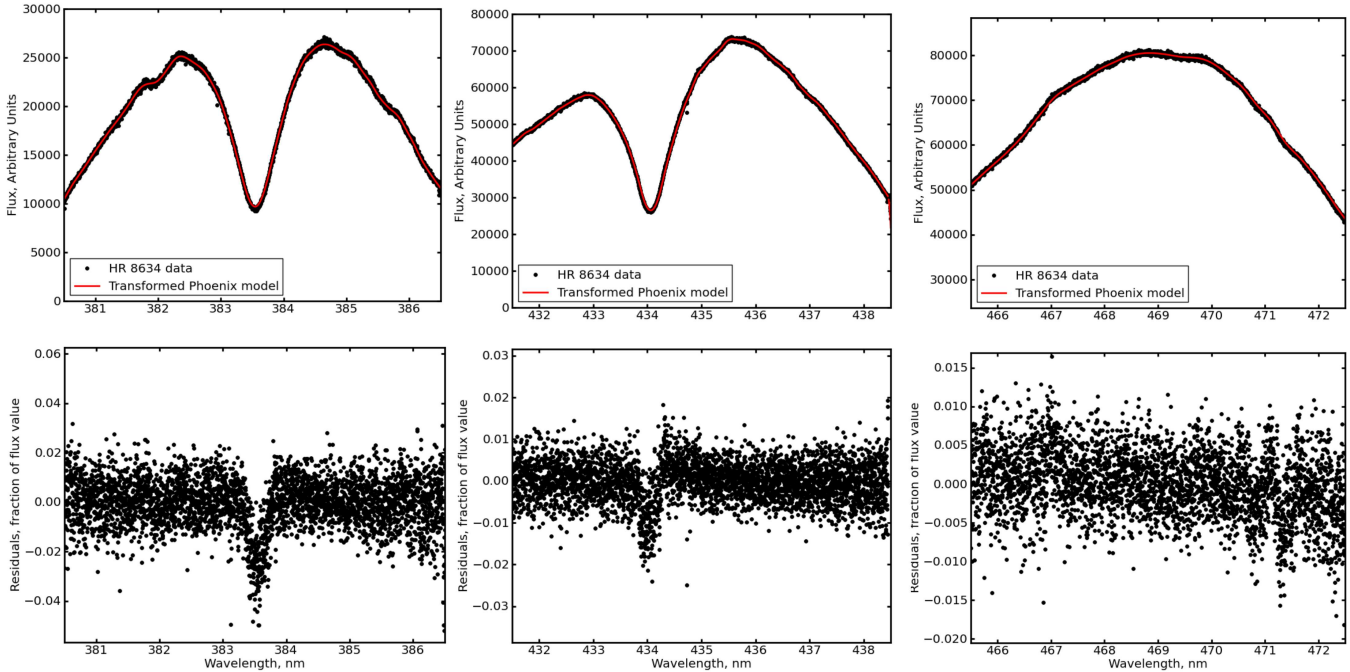} 
   \caption{ Three representative orders from a spectrum of HR\,8634, spanning roughly 80 angstroms each.  Data points and residuals in black with the best fit transformed PHOENIX model in red. The bottom panel shows the residuals in fractional values of the total flux at each pixel. The trend in residuals around the lines are due to the imperfect fit of the theoretical line profile model to the data. The fit precision is worse for theoretical models than for relative radial velocities, which is part of why the absolute radial velocity precision is worse than the relative radial velocity precision.}
   \label{absfitting}
\end{figure*}

In addition to measuring relative radial velocities for the stars in our sample, we also measured absolute radial velocities for these stars using a somewhat modified version of our technique. 
Previous groups have made use of CPS spectra for measuring absolute radial velocities: for example, \cite{2012arXiv1207.6212C} analyzed over 29,000 spectra of 2046 F-, G-, K-, and M-type stars \citep[see also][]{2002ApJS..141..503N}. In this work, we analyzed an additional $\sim$3000 spectra of 213 more massive A- and B-type stars that were not included in \citet{2012arXiv1207.6212C}. 

We measured the absolute radial velocity for each star in our sample using the same algorithms described in Section \ref{methody}. However, instead of using a spectrum of the star itself as a template, we performed the fit using a PHOENIX model stellar spectrum of a hot A- or B-type star, using model:

\begin{equation}
\textcolor{black}{
\begin{split}
\mathcal{I}_{m}(i, n) = \rm{F}(i,n) &\cdot S_m \left[\lambda_{\rm{extrap}}(i, n)\left(1+{\frac{V_{\rm Dop}}{c}}\right) \right] \\ &* G[V_{\rm{rot}}\sin{I}, R]
\end{split}
}
\end{equation}

\noindent which is similar in form to Equation \ref{model}: \rm{F}(i,n) is the continuum level of the datum spectrum, $S_{m}$ is the unperturbed flux level of the PHOENIX spectrum model, $I$ is the inclination of the stellar spin axis with respect to the line-of-sight, $V_{\rm{rot}}$ is the equatorial stellar rotation velocity, the ``$*$'' symbol denotes a convolution, and $G[V_{rot}\sin{I}, R]$ is the broadening kernel, defined by:

\begin{equation}
\begin{split}
G[V_{rot}\sin{I}, R] = &\left(\frac{2(1-\epsilon)}{ \pi (1-\epsilon/2)}\right) \left(1 - (\Delta\lambda /\Delta\lambda_{L})^{2}\right)^{0.5} \\& +\left(\frac{\pi \epsilon}{2 \pi (1- \epsilon/3)} \right) (1 - (\Delta\lambda /\Delta\lambda_{L})^{2})
\label{kkernel}
\end{split}
\end{equation}
\noindent when $R$ is resolution, in units of $\delta\lambda$ per pixel; $\Delta\lambda /\Delta\lambda_{L}$ is a unitless argument, bounded by -1 and 1 describing the position on the star at which the kernel is to be evaluated; $\epsilon$ is the limb-darkening coefficient \citep[taken to be $\epsilon = 0.6$ in our analysis,][]{1976oasp.book.....G}.

The PHOENIX atmospheric models have been developed over the last fifteen years for modeling the spectra of wide range of stellar masses and spectral types \citep{1999ApJ...512..377H, 2010ascl.soft10056B}. The model spectra used in this analysis have metallicities as given in \cite{2009ARA&A..47..481A}. The models were generated from effective stellar temperatures as available in the literature for each individual star, ignoring metallicity and surface gravity variations: our fits do not derive stellar qualities beyond the doppler shift, and thus all we require from a model is the best possible template for deep, broad emission lines.
Although using the PHOENIX spectrum as the model requires significantly more exploration through parameter-space to find the Doppler shift, continuum shape, and line-broadening parameters, the result is a description of the absolute motion of the star. 
We applied a barycentric correction and a slight correction to match the zero-point set by IAU standard stars \citep[e.g.][]{2002ApJS..141..503N, 2012arXiv1207.6212C, 2014AJ....147...39C}. We do not correct for the gravitational redshift due to either the host star or our own sun, as those effects would be on the order of a few \ms, which is well below our target precision \citep{2014PASP..126..838W}. 

Although our method is not optimized to find $V \sin{I}$, this quantity is a byproduct of our fits and so we present derived values of $V \sin{I}$ and its error, $\sigma_{V \sin{I}}$, for each star in our sample in Table \ref{absrvs}, assuming zero stellar turbulence.

A final fit to a model and residuals are shown in Figure \ref{absfitting}.  As evident in the figure, there is some discrepancy between the PHOENIX model atmosphere and the observations. Much of the mismatch is in the line core. However, this is not a major concern given that the Doppler content is contained in the line wings. %In the line cores,  $dI/d\lambda \approx 0$, and the mismatch between model and observation has little effect on our Doppler measurements.

\section{Results}

\label{highscatter}

We applied our radial velocity measurement technique to observations of our sample of 213 hot stars. To narrow this sample to contain only those potentially in binary stars, we selected a sub-sample containing stars that had been observed more than seven times with more than three epochs of observations and additionally showed a large ($> 3 \sigma$, when $\sigma \approx$ 1 \kms\ is our method precision) amount of scatter in their relative radial velocities. Here, epoch refers to the night upon which the observation was taken (such that if many spectra were taken in a single night, it would only be one epoch of observations). This sub-sample contains 13 stars that fit these criteria. Of these, one (HR\, 1178) was excluded from the sample due to being a blended binary as identified in literature, two were excluded due to a significant number of their observations being taken during twilight, leaving 10 stars for further analysis. Of these, three are in detectable multiple systems. We describe our measurements for each of these stars individually in Section \ref{multiple}. We also extracted radial velocities for all 213 stars and report absolute radial velocities in Table \ref{absrvs_table}.

Many observations were taken sequentially within the same night, often in clusters of three observations within $\sim5$~minutes. To calculate an appropriate bin size for these points, we found the upper limit of semi-amplitude that could occur due to a single-lined companion at the smallest orbital radii possible. We found it was possible to bin data in 4 minute intervals for A-type stars and 15 minute intervals for B-type stars without risk of missing extremely short-period binaries. 

\subsection{Measured RV Precision}

We used the radial velocity measurements of the stars with extended time series to both estimate the radial velocity precision of our method, and set limits on any radial velocity trends. To do this, we fit the measured radial velocities to a linear model, including a term for radial velocity jitter, using a Markov Chain Monte Carlo (MCMC) algorithm with an affine invariant ensemble sampler \citep[adapted for IDL from the algorithm of][]{goodman, 2013PASP..125..306F}. Our best-fit radial velocity trends and jitter are reported in Table \ref{trends}.  

\begin{deluxetable*}{lccc}
\tablewidth{0pt}
\tablecaption{ Best--Fit Linear Trends and RV Jitter \label{trends}}
\tablehead{
\colhead{Star} & \colhead{Trend (km~s$^{-1}$~yr$^{-1}$)} &\colhead{Significance ($\sigma$)} &  \colhead{Jitter (km~s$^{-1}$)}}
\startdata
HR\,1679& $-0.16 \pm 4.38$& 0.04&$7.3   \pm 4.3$\\
HR\,2845& $0.33 \pm 0.43$& 0.77&$2.0   \pm 0.6$\\
HR\,3799& $-0.04 \pm 2.09$& 0.02&$2.3  \pm1.0$\\
HR\,4468& $-0.33 \pm 0.23$& 1.41&$0.9  \pm 0.2$\\
HR\,5511& $0.09 \pm 0.16$& 0.54&$1.5   \pm 0.3$\\
HR\,5849& $0.10 \pm 0.35$& 0.29&$1.0   \pm 0.3$\\
HR\,5867& $0.96 \pm 0.21$& 4.52&$0.5   \pm 0.2$\\
HR\,7708& $-0.36 \pm 2.59$& 0.14&$13.4 \pm 2.7$\\
HR\,8028& $-1.58 \pm 0.29$& 5.51&$1.3  \pm 0.3$\\

\enddata
\tablecomments{Significance $\sigma$ refers to the magnitude of the trend divided by its uncertainty. Stars with too few ($<7$) data points are excluded from this table. The best--fit jitter values indicate our precision is typically 1-2 \kms. Note that HR\,1679 and HR\,7708 have higher levels of scatter than is typical of this technique, possibly indicating close binary companions. \textcolor{black}{HR\, 3067, discussed in section \ref{thirtysixtyseven} is excluded from this table.}}
\end{deluxetable*}

We find that typically, for stars without known binary companions, the best-fit jitter is between 0.5 and 2 km/s, which we take as the typical precision of our technique. Our data for some stars without known binary companions are consistent with higher values of jitter because the stars have fewer measurements to constrain jitter. Finally, we detect two significant radial velocity trends, which we discuss further in Section \ref{multiple}
 
\subsection{Absolute Radial Velocities} 

We report our measurements of absolute radial velocities for all 213 stars in our sample in Table \ref{absrvs_table}. Table \ref{absrvs_table} lists the average absolute radial velocity over all observations of each star in our sample, as well as the number of observations and the time baseline of all observations. We report the time baseline of the observations to help avoid contamination with spectroscopic binaries --  With a long enough baseline of observations it is possible to separate the overall motion of the target from the periodic motion due to companions, and we indeed see some stars with obvious Keplerian motion. However, without a sufficiently long baseline, it is unclear whether an observed radial velocity is due to the star's absolute motion or if it contains an instantaneous snapshot of a star's motion due to the effect of a companion. 
Only a subset of the A- and B-type calibrator stars were observed a sufficient number of times over a long enough time baseline to make an informed statement on their radial velocities over time. 
For this reason, the $T_{\rm baseline}$ in days is included in Table \ref{absrvs_table}, to provide a context for each absolute radial velocity measurement.
\begin{figure}[htbp] %  figure placement: here, top, bottom, or page
   \centering
   \includegraphics[width=3.4in]{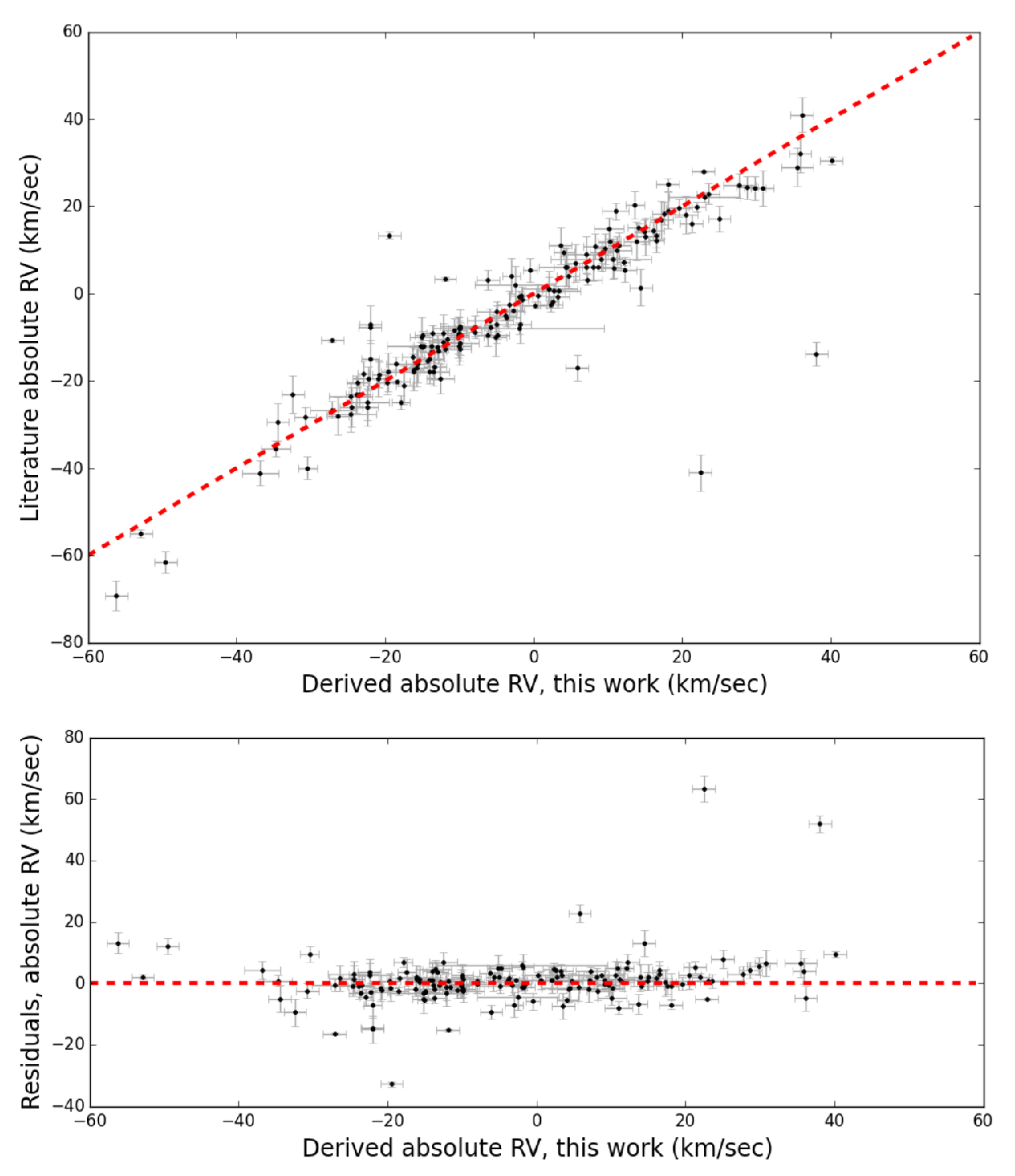} 
   \caption{ \textcolor{black}{Top panel, a plot of the absolute radial velocity values derived in this work against literature values \citep[drawn from][]{2006AstL...32..759G}. Bottom panel, residuals between our values and literature values}. The RMS Error is 8.63 /kms. }
   \label{comparervs}
\end{figure}

Both \cite{2012arXiv1207.6212C} and \cite{2002ApJS..141..503N} were able to calculate absolute radial velocities for F-, G-, K-, and M-type stars to a precision of roughly 0.1 \kms. We find that our method yields a median precision of 1.5 \kms\ (with can be further delineated into a best case scenario precision as good as 0.5 \kms\ for low-mass, A-type stars, and a worst-case scenario of 2 \kms\ for massive, rapidly rotating B-type stars). \textcolor{black}{In Figure \ref{comparervs}, we plot a schematic comparison between our derived values and the compilation presented in \cite{2006AstL...32..759G}. }We adopt 1.5 \kms\ as the typical uncertainty for our absolute measurements, which is somewhat higher than our errors for relative radial velocities, due to discrepancies between the PHOENIX models and the observed spectra.

\subsection{\textcolor{black}{Binary Systems}}
\label{multiple}
In our sample of 213 stars, each star has an average of 13 spectra covering 3 epochs. With such sparse temporal sampling, it is difficult to find true periods and fit orbits. 
Lomb-Scargle analysis of the radial velocity time series for each target often finds spurious short-period signals, due to aliasing, which is particularly troublesome for sparsely sampled targets \citep{2010ApJ...722..937D}.
For this reason, we rely on the scatter of the radial velocities compared to the measurement uncertainties as a simple indicator of the potential presence of a companion, and then consider the data in the context of what is already present in the literature about these sources. Using literature periods as starting points, we are able to confirm, and, in some cases, refine, what has been reported about the binarity of these massive, bright stars.  

Many of the targets in our sample are previously studied binaries \citep{2012MNRAS.424.1925C}.
We observe one blended double-lined binary, HR\,1178 \citep{1965ApJ...142.1604A, 2004A&A...425L..45Z}, which we exclude from our sample because our modeling technique does not account for multiple lines in the spectra. \textcolor{black}{Our method is optimized to find single-lined binaries, and is not presently capable of dealing with double-lined binaries. }

We additionally exclude sources with high scatter but insufficient phase coverage, that is, observations at fewer than seven epochs.
Out of the 10 stars with at least seven observations, we identify two previously unknown binary systems. We additionally detect the stellar companion to HR\,3067, previously found using methods other than radial velocities.
We summarize literature and our observations for each of these systems.

\subsubsection{HR\,3067}
\label{thirtysixtyseven}
\begin{figure}[htbp] 
   \centering
   \includegraphics[width=3.5in]{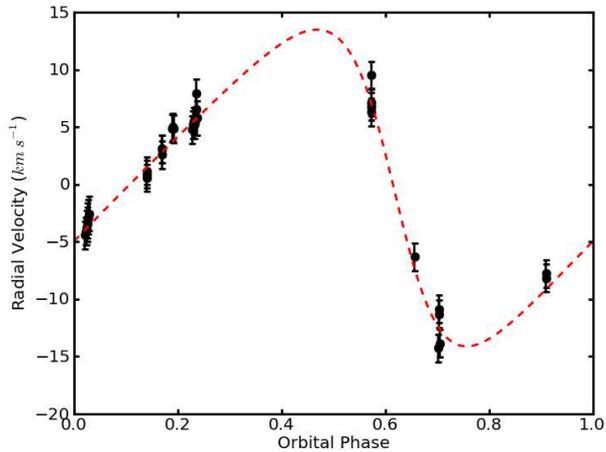}
   \caption{The phase-folded radial velocity time series of HR\,3067, an A-type star with a predicted companion at an orbital period of 1.5 years \citep{2012A&A...546A..69M}. The RVs show an orbit with a longer period than astrometry predicted and significant eccentricity. }
   \label{HR3067}
\end{figure}

HR\,3067 is a bright star with spectral type A3, and was observed by CPS 23 times over a time baseline of 5 years. Astrometric observations from \Hipparcos\ identified a companion with an orbital period of 1.59 years, and fit the data with a zero--eccentricity orbital model. To our knowledge, no radial velocity confirmation of this companion or measurement of the eccentricity exists in the literature. In our data, we indeed detect significant radial velocity variations \textcolor{black}{(a scatter roughly 10 times the measured scatter of the method; see Table \ref{deluxta})}, and searched for periodic signals with a Lomb-Scargle analysis. The periodogram analysis of the radial velocity time series of HR\,3067 finds several potential short-period peaks; however, due to poor sampling and aliasing \citep[e.g.][]{2010ApJ...722..937D}, we cannot uniquely determine the true period. 

We fit a Keplerian model to HR\,3067's radial velocity time series. We adopted the astrometrically derived period as a starting point for the fit, and fit the radial velocity time series to a Keplerian using the IDL program \texttt{rvlin} \citep{2009ApJS..182..205W, 2012ascl.soft10031W}. The result is plotted in Figure \ref{HR3067}.
To determine errors on these parameters, we used a bootstrap Monte Carlo method, as used in \cite{2007ApJ...665..785J}. This is done by subtracting a model generated from the best-fit Keplerian from the measured radial velocities, then computing the residuals between the two. The residuals are then randomly reassigned to data points, and \texttt{rvlin} used again to fit a new best-fit Keplerian. The mean and standard deviation of a distribution composed of 1000 such trials was adopted as the system parameters and uncertainties.

We estimated HR\,3067's mass using the online Padova model interpolator\footnote{\url{http://stev.oapd.inaf.it/cgi-bin/param}}, which uses the technique of \citet{2006A&A...458..609D} to calculate masses based on photometry, parallaxes and spectroscopic parameters. Assuming HR\,3067's observed spectral type of A3, a temperature of roughly 8750 K and solar metallicity, we estimate that it has a mass of roughly 2.2 $M_{\odot}$. Given the primary's mass of 2.2 $M_{\odot}$ and the period derived from our RVs, the binary mass function of $0.15 M_{\odot}$ translates to a minimum mass for the secondary component of about $1 M_{\odot}$. 
Additionally, hints of the spectral lines of the secondary component are visible in the HIRES spectra. Follow-up observations and analysis are ongoing, and the result of further analysis will be presented in a future work.

We show our phase folded radial velocity measurements along with our best--fit model in Figure \ref{HR3067}. The best-fit orbital parameters are provided in Table \ref{3067params}.

\begin{deluxetable}{lcr}
\tablewidth{0pt}
\tablecaption{ Radial velocity time series for HR\,3067 \label{3067data}}
\tablehead{
\colhead{Date (JD)} &\colhead{Relative RV} & \colhead{Uncertainty} }
\startdata

2454808.016881 & -7.99011456 &0.906307 \\
2454963.801250 & 0.84506144 &0.851385 \\
2454983.734892 & 2.89520244 &0.730252 \\
2455255.838258 &7.40614944 &1.730396 \\
2455311.718924 & -6.35259256 &0.912301 \\
2455342.728611 &-14.29464556 &1.012084 \\
2455343.731059 &-11.11869856 &0.923183 \\
2455344.732477 &-13.89147756 &0.842331 \\
2455671.717002 &  4.87441244 &1.175328 \\
2455672.718090 &4.98378644 &1.273581 \\
2455673.721030 &4.81967244 &1.293804 \\
2455697.725914 &4.73993644 &1.377920 \\
2455698.726088 &5.18161544 &1.201775 \\
2455699.725868 &5.49505144 &1.187538 \\
2455700.725764 &5.18047444 &1.042351 \\
2455702.741458 &7.22616644 &0.986378 \\
2456907.153611 &-5.3110145 &1.071243\\
2456908.154329 &-5.6374542 &1.141087\\
2456909.154225 &-5.9577133 &1.184617\\
2456910.146331 &-4.2697303 &0.918832\\
2456911.145278 &-4.5781529 &1.943838\\
2456912.153484 &-4.8835522 &1.307474\\
2456913.153634 &-3.1805621 &1.234314\\
\enddata
\label{deluxta}
\end{deluxetable}

\begin{deluxetable}{lcr}
\tablewidth{0pt}
\tablecaption{ Best-fit orbital parameters for HR\,3067 \label{3067params}}
\tablehead{
\colhead{Orbital Parameter} &\colhead{Value} & \colhead{Uncertainty} }
\startdata
$p$       & 674.64 days & 7.35 days \\
$e$       & 0.38  & 0.05 \\
$\omega$  & 93.91$^{o}$  & 3.12$^{o}$ \\
$k$       & 13.80 km~s$^{-1}$   &  0.79 km~s$^{-1}$ \\

\enddata

\end{deluxetable}

\subsubsection{HR\,5867}
HR\,5867 is an A3-type star that has been studied in the past and found to have many different possible companions, of varying separations. 
\cite{1929PA.....37...16V} identified HR\,5867 as a quadruple system, the components of which were separated by 30" from the primary.  
\cite{2011ApJS..192....2S} reported a very wide companion to HR\,5867. 
\cite{2014MNRAS.437.1216D} identified a further companion at 1643.04 arcseconds.

We find a linear trend in our new radial velocities, suggesting the presence of a close-in, previously unstudied companion. Our new radial velocities span a baseline of more than 1000 days, but do not catch a turn-over in the radial velocity curve. As shown in Figure \ref{5867trend}, an MCMC fit to a linear model finds a significant slope, indicating long-term motion in the star. This companion has not been previously identified. 
Since we have a measure of the radial velocity trend as well as an astrometric measure of the separation between the primary and visible companions, we can determine the companion mass using \cite{2014ApJ...785..126K}: 
\begin{equation}
M_{\rm comp}/M_\Sun = 5.34 \times 10^{-6}  \left(\frac{d}{\rm{pc}} \ \frac{p}{\rm{arcsec}}\right)^{2} \left| \frac{d(RV)}{dt} \right| \Phi
\label{knutson}
\end{equation}
when $d$ is distance to the system, $p$ is separation in arcseconds between the primary and companion, $\frac{d(RV)}{dt}$ is the radial velocity trend, and $\Phi$ is a function of the  inclination angle, eccentricity, longitude of periastron, and phase in orbit, which assumes a minimum value of $\sqrt{27}/2$ \citep{1999PASP..111..169T, 2002ApJ...571..519L, 2014ApJ...785..126K}.
Using Equation \ref{knutson}, \cite{2007A&A...474..653V}'s measured parallax of 21.03 $mas$, and the separations of the previously measured companions \citep{1929PA.....37...16V,2011ApJS..192....2S, 2014MNRAS.437.1216D}, and our new value of $\frac{d(RV)}{dt} = 0.96$ km/sec/year, we find that if the RV trend we observe was caused by one of the known visual binary companions to HR\,5867, the companion would have a minimum mass of $2\times10^{4} M_{\odot}$. This result is unphysical, so we conclude that the companion inciting the radial velocity trend that we see must be an undiscovered companion. 

Estimating HR\,5867 as a 2.2 $M_{\odot}$ star using the Padova interpolator described above, we can estimate the mass of the companion using \citep{2007ApJ...657..533W}:
\begin{equation}
\frac{m^3 \sin^{3}{i}}{(m + M_{*})^{2}} = \frac{P K^{3} (1-e^{2})^{3/2}}{2 \pi G}
\label{wright}
\end{equation}
where $m$ is the mass of the companion, $M_{*}$ is the mass of the primary, $P$ is the period of the orbit, $K$ is the amplitude of the radial velocity signal, and $e$ is the eccentricity. Using the minimum amplitude for HR\,5867 ($K \geq 3.15$ \kms), the minimum period ($P \geq 6.3$ years), and assuming that $e=0$ and $M_{*} = 2.2 M_{\odot}$ and $i = 90\degree$, we find that the companion must have a minimum mass of $m \geq 0.37 M_{\odot}$. 
This could be further constrained with additional radial velocity measurements, particularly if the new data cover an inflection point.

\subsubsection{HR\,8028}
HR\,8028 is a A1-type, main sequence star with a speckle companion roughly 0.1" away \citep{2012AJ....143...10H}. The astrometric separation has been measured several times: these values are presented in Table \ref{7028}, along with the instantaneous separations derived using HR\,8028's parallax, which is 8.71 mas $\pm$ 0.34 mas \citep{2007A&A...474..653V}, corresponding to a distance of 114 $\pm$ 4 parsecs. 

\begin{figure}[htbp] %  figure placement: here, top, bottom, or page
   \centering
   \includegraphics[width=3.5in]{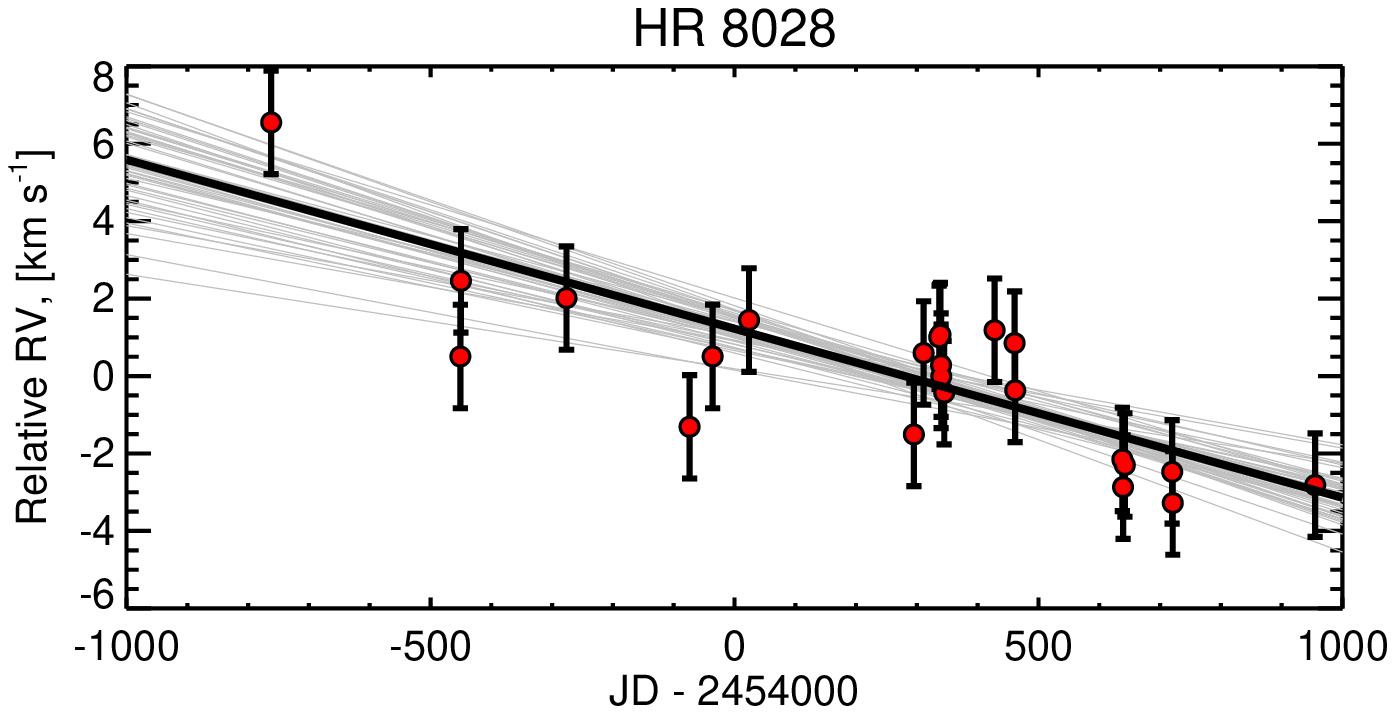}
   \includegraphics[width=3.5in]{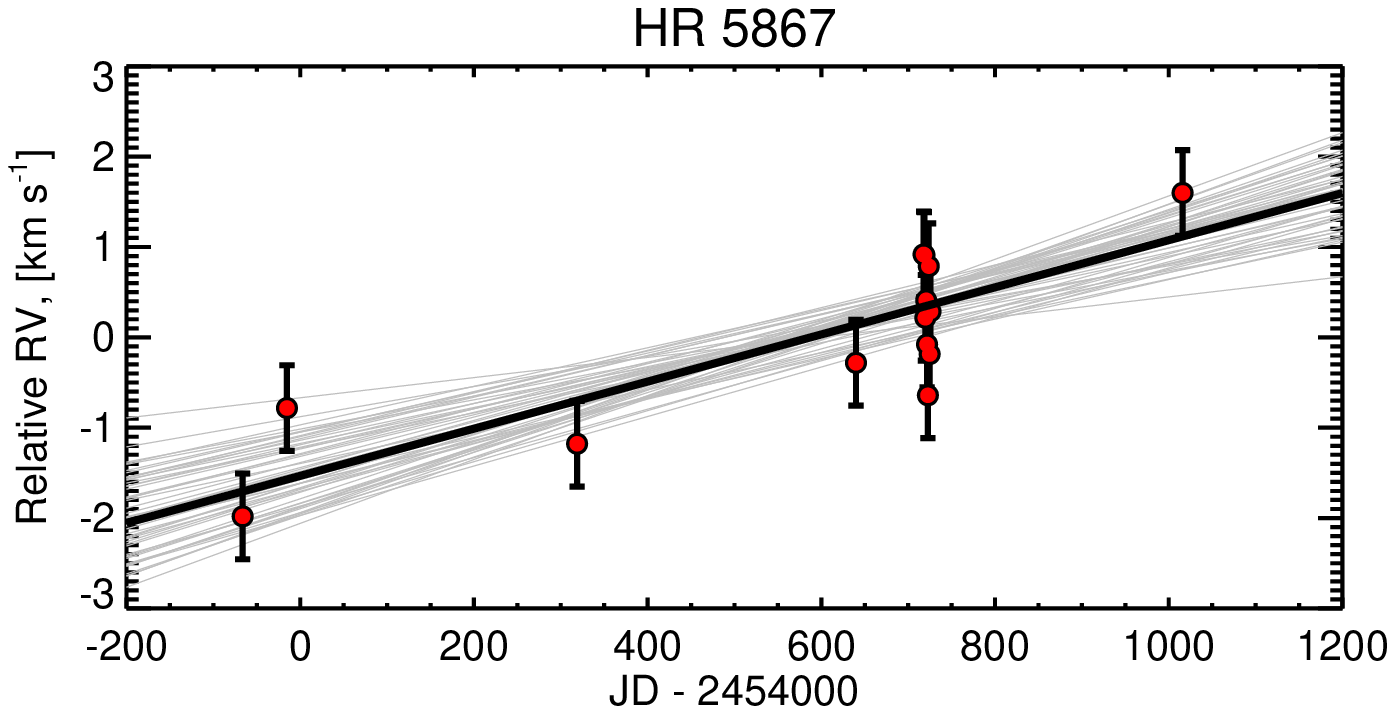}
   \caption{Top: A linear fit to the radial velocity points for HR\,5867 excludes a zero slope (no trend) to 4.5$\sigma$. Grey lines show random draws from the MCMC posterior. This indicates the presence of a companion forcing the trend, inducing an amplitude variation of 0.96 \kms yr$^{-1}$. Bottom: Same as top, but for HR\,8028. We attribute this trend to a companion detected in speckle imaging of this star. The best--fit trend is -1.58 \kms yr$^{-1}$, and is significant at the 5.5$\sigma$ level.}
   \label{5867trend}
\end{figure}

\begin{deluxetable*}{lccccc}[h!]
\tablewidth{0pt}
\tablecaption{ Astrometric Measurements for HR\,8028 \label{7028}}
\tablehead{
\colhead{Date} &\colhead{\textcolor{black}{$\rho$, Ang.} Sep (arcsec)} &  \colhead{Position Angle (\degree)} & \colhead{\textcolor{black}{Separation(AU)}} &  \colhead{Reference} }
\startdata

1989.7114 & 0.262  & 89.7114 & \textcolor{black}{29.9} &\cite{1999AJ....117.1890M}  \\
2003.5383 & 0.084  & 204.6 & \textcolor{black}{9.6} &\cite{2008AJ....136..312H}  \\
2003.5384 & 0.0849 & 202.1 & \textcolor{black}{9.6} &\cite{2008AJ....136..312H}  \\
2008.4722 & 0.113  & 169.8 & \textcolor{black}{12.9} &\cite{2012AJ....143...10H} \\
2009.4498 & 0.140  & 177.5 & \textcolor{black}{16} &\cite{2012AJ....143...10H} \\
2009.4498 & 0.144  & 175.0 & \textcolor{black}{16.4} &\cite{2012AJ....143...10H} \\
2009.4578 & 0.137  & 176.2 & \textcolor{black}{15.6} &\cite{2012AJ....143...10H} \\
\enddata
\end{deluxetable*}
As shown in Figure \ref{5867trend}, an MCMC fit to a linear model finds a significant slope, indicating long-term motion in the star. The companion inciting this motion must have a minimum period of 10 years, consistent with the orbital separations found via astrometry by \cite{1999AJ....117.1890M}, \cite{2008AJ....136..312H}, and \cite{2012AJ....143...10H}. The $\frac{dv}{dt} = 1.58$ \kms yr$^{-1}$ and separations in Table \ref{7028} allows us to calculate a minimum dynamical mass for the companion based on the local radial velocity slope, assuming a mass of 2.3 $M_{\odot}$ (from the Padova interpolator) for the A1 primary.
Using Equation \ref{knutson} and the separations summarized in Table \ref{7028}, we find the mass of the companion to be a minimum 2 $M_{\odot}$, which would make the companion of comparable size to the primary. This is unlikely, considering the magnitude difference between the primary and secondary evident in speckle imaging \citep[2.1 magnitudes,][]{1999AJ....117.1890M}. The speckle companion seems to be moving relative to the primary (roughly 0.3 arcsec in 20 years) at a rate comparable to the proper motion of the primary (0.5 arcsec in 20 years). It is possible that the speckle companion is a background star, and the companion we detect in radial velocities is different. 
\textcolor{black}{Using Equation \ref{wright} with lower bounds of 8 years for the period (due to unconstrained eccentricity, we cannot exclude a 8 year period for an eccentric orbit) and 5 \kms\ for the radial velocity amplitude yields a minimum mass companion of $m \ge 0.7 M_{\odot}$. The period and amplitude both have only lower limits due to the lack of turn-over in the radial velocity curve. }
 Further refinement of the period of HR\,8028's companion (and thus lower bounds on the its mass) will be possible once a turnover is measured in the radial velocity measurements. 

\section{Discussion}

In this work, we present data on some of the closest, brightest stars in the sky. Even though these stars have been studied for over a century, it is still possible to make discoveries by using existing data in new ways. 

Some of our absolute radial velocity measurements are new and have not yet been presented in literature, while many others serve as an update to previous literature values. These new measurements will serve as additional reference measurements for programs studying the kinematics of bulk stellar flow. The updated absolute RVs can also be combined with with long-term measurements of these stars to extend the time baseline of RV monitoring of these sources. Moreover, additional absolute RV measurements will help the Hundred-Thousand-Proper-Motion survey. \cite{2012A&A...546A..61D} found that many sources, including some in our sample, need additional radial velocity measurements to be useful to the Hundred-Thousand-Proper-Motion survey. 

Calibrator spectra are an underutilized resource with the potential to do new science. Our method of fitting radial velocities for massive stars could be applied to spectrographs other than HIRES, allowing for more radial velocity measurements of various types of rapidly rotating stars. A compilation of calibration spectra from other spectrographs could significantly increase the size of our dataset and help the sampling issues that prevented us from identifying more spectroscopic binaries. Infrared spectrographs in particular could substantially increase the number of calibration spectra, because taking spectra of rapidly rotating hot stars to calibrate spectral features from Earth's atmosphere is common practice among infrared astronomers \citep[e.g.][]{2013ApJ...767..111M}.

Additionally, though this method was developed for use on A- and B-type stars, it can be used for any target with broadened spectral features. A modified version of this method was used in \cite{2013ApJ...767..111M} to find radial velocities for a rotating M3-dwarf with a $V_{\rm rot} \sin{i} = 19.67 \pm 0.52$~\kms. 
This method could be used for observations of young stars as well as those of massive stars.

\section{Summary}
We have developed a method to extract radial velocity measurements from A- and B-type stellar spectra using a forward modeling approach that simultaneously fits the star's radial velocity with the echelle spectrograph's blaze function. Our technique utilizes an extrapolation of the wavelength solution from the iodine calibration region and derives a radial velocity measurement from the entire spectrum simultaneously. 
This method makes use of qualities of the spectra that are usually weaknesses in radial velocity work, namely the broad spectral lines and multiple featureless orders, to instead serve as strengths in the fitting process. Fitting RVs in this non-traditional manner allows for the analysis of echelle spectra of rapidly rotating stars, which cannot be processed with traditional pipelines. 

We found that with our technique, we attain a precision of 1.0 \kms (0.5 \kms - 2.0 \kms) for relative radial velocities. For absolute radial velocities, which rely upon PHOENIX stellar model spectra as radial velocity templates, the precision is a bit worse (1.5 \kms). 

We detect several sources with a high degree in scatter between successive radial velocity measurements taken over the course of anywhere from one to six years. Since these stars were observed as calibrators and not science targets, they are often significantly under-sampled. The sparse sampling of each individual target limits our ability to totally characterize these detections. We also detect two significant long--term radial velocity trends (HR\,5867 and HR\,8028), and redetect a previously known astrometric binary (HR\,3067).  Two detections have radial velocity time series with slopes that cannot be attributed to any currently known companion in the system (HR\,5867) or nearby speckle star (HR\,8028). Though we only see an unknown fraction of the phase in our time series, we compute the minimum masses of these new potential companions to be 0.37 and 0.70 $M_{\odot}$, respectively.

\acknowledgements
We thank Emily Rauscher for her careful review of the manuscript and helpful suggestions. J.B. thanks Philip Muirhead for useful conversations and Iryna Butsky for useful comments on the manuscript. \textcolor{black}{We thank the referee, Davide Gandolfi, for his extremely helpful suggestions that led to a vastly improved paper, as well as his suggestions for future directions to take this work. }
J.B. and A.V are supported by the National Science Foundation Graduate Research Fellowship, Grants No. DGE 1256260  and DGE 1144152, respectively. J.B. would like to thank Mr. and Mrs. Kenneth Adelman for providing funding for her 2012 Alain Porter Memorial SURF Fellowship, during which this research was begun. J.A.J is supported by generous grants from the David and Lucile Packard and Alfred P. Sloan Foundations.
This research has made use of NASA's Astrophysics Data System, the SIMBAD database and VizieR catalog access tool, operated at CDS, Strasbourg, France. The data presented herein were obtained at the W.M. Keck Observatory, which is operated as a scientific partnership among the California Institute of Technology, the University of California and the National Aeronautics and Space Administration. The Observatory was made possible by the generous financial support of the W.M. Keck Foundation. The authors wish to recognize and acknowledge the very significant cultural role and reverence that the summit of Mauna Kea has always had within the indigenous Hawaiian community.  We are most fortunate to have the opportunity to conduct observations from this mountain.

\clearpage
\appendix
\section{Expected precision} \label{precision}

The theoretical best precision on a radial velocity measurement depends on the signal--to--noise ratio of the measurement, the typical width and depth of spectral features, and the number of spectral features used to calculate the Doppler shift. \citet{1996PASP..108..500B} derives the expected radial velocity precision:
\begin{equation}
\sigma_{V} = \left[ N_{\rm line}  \sum_{i} \left( \frac{d\mathcal{I}_{i} / dV}{\epsilon_{i}} \right)^{2} \right]^{-1/2} \approx \frac{1}{S\sqrt{N_{\rm pix} N_{\rm line}}} \frac{\Delta V}{\Delta \mathcal{I}}
\end{equation}

when $\epsilon_{i} = \frac{\sqrt{N_{\rm{photons}}}}{N_{\rm{photons}}}$, $N_{lines}$ is the number of spectral lines, $N_{pix}$ is the number of pixels across which each line occurs, $\Delta\mathcal{I}$ is the relative intensity depth of the spectral features, $\Delta V$ is the average range in wavelength across which this intensity depth occurs, and $S$ is the signal to noise ratio. 

For a solar-type spectrum, a typical line might encompass six pixels, with an overall line width of $dV = 2.5$ \kms\ and a relative intensity depth of $d\mathcal{I} = 0.2$. With a typical observation SNR of 200 \textcolor{black}{and $N_{lines}$ = 100 spectral lines,} this leads to an expected radial velocity precision of 3.6 \ms\ \cite{1996PASP..108..500B}. 

To calculate the best-case precision for our method, we calculate the precision for a star in our sample with the median amount of broadening. For such a star, we estimate the relative intensity depth to be roughly equivalent to that for a low-mass star, $d\mathcal{I} = 0.2$ (see Figure \ref{comparison} for a visualization of the comparison). The average line in such a star achieves this intensity dip over ${\Delta}V = 800$ \kms and $N_{\rm{pix}} = 500$ pixels, and has $N_{lines}=10$ of these lines. Using these median parameters and a standard SNR of 100, we find that were we to be photon limited, we could expect our average precision to be roughly $0.6$ \kms.

\clearpage
\LongTables
\begin{deluxetable*}{llllllll}
\tablewidth{0pt}
\tabletypesize{\scriptsize}
\tablecaption{\label{absrvs_table} Summary of all observations}
\tablehead{
  \colhead{Star} & 
  \colhead{Abs. RV (\kms)}     &
  \colhead{$\sigma_{RV}$ (\kms)} &
  \colhead{$V\sin(I)$ (\kms)}     &
  \colhead{$\sigma_{V\sin(I)}$ (\kms)} &
  \colhead{$N_{obs}$}     &
  \colhead{$T_{baseline}$ (days)}  & 
  \colhead{Notes} 
}
\startdata
HR10          & -10.92           & 1.5            & 252 & 17      & 6                  & 462.677778            &    c   \\
HR1002        & -10.06           & 1.00     & 132 & 6             & 13                 & 1098.160023             &       \\
HR1062        & 14.11            & 1.5           & 120 & 13       & 7                  & 175.637489              &    c   \\
HR1087        & -1.67            & 1.5             & 190 & 14     & 10                 & 0.018322                &   d    \\
HR1239        & 16.52            & 0.40         & 38 & 11        & 16                 & 1834.113172             &       \\
HR1260        & 5.82            & 1.5              & 168 & 23    & 6                  & 0.089051                &    c, d   \\
HR1261        & 7.11             & 0.91              & 251 & 15     & 12                 & 392.0197                &       \\
HR1273        & 4.07             & 1.39            & 55 & 9     & 10                 & 767.998484              &       \\
HR128         & -21.99           & 1.17            & 194 & 13      & 9                  & 434.995069              &       \\
HR1289        & -22.01           & 1.5           & 265 & 10        & 3                  & 0.001412                &   c, d    \\
HR15          & -6.35            & 1.5           &  95 & 40       & 3                  & 1236.918252             &   d    \\ 
HR1500        & 12.24             & 1.55        &  249    & 23       & 12                 & 69.792176               &       \\
HR1544        & 28.70            & 1.5          & 228 & 19        & 7                  & 1097.996562             &    d   \\
HR1567 & 30.78   & 1.5    &  77  & 27 & 3   & 0.087615    & c, d \\
HR1574 & 10.12   & 1.97  &   155 & 14  & 15  & 86.672651   &  \\
HR1621 & 29.78   & 1.5    &   312  & 26  & 6   & 447.719641  & c  \\
HR1641 & 12.09   & 0.85  &    115  & 15   & 35  & 2361.619584 & \\
HR1679        & -2.51            & 5.43            &  327 & 24      & 37                 & 978.285498              &   a    \\
HR1786        & -10.09           & 1.5             & 190  & 19      & 6                  & 0.002546                &   c, d    \\
HR1789        & 19.98            & 1.5             &  287   &  24    & 6                  & 28.939711               &    c, d   \\
HR179         & -12.58           & 1.75             &  134   & 8    & 11                 & 1236.393646             &       \\
HR1806 & 17.51   & 1.5       &     202  & 41       & 3   & 0.001783    &  c, d\\
HR1858 & 36.12   & 1.5    &     199  &  26      & 3   & 0.001342    &  c, d\\
HR1873 & 5.65    & 1.5    &     231  &   29     & 3   & 0.001632    &  c, d\\
HR193         & -14.93           & 1.5               &  221     &     16        & 7                  & 688.042025              &  c     \\
HR2155 & 35.89   & 1.5    &      234 &   15     & 3   & 0.00169     &  c, d\\
HR2198        & 19.54            & 1.72                 &    248   &     8      & 14                 & 507.819283              &       \\
HR2209 & -13.26  & 1.5    &    245   &   31     & 3   & 0.001412    & c, d \\
HR223         & 7.01             & 2.97            &   115    &       13         & 21                 & 765.087974              &       \\
HR2231 & 22.83   & 1.5   &     244  &    22     & 5   & 1.993738    & c, d \\
HR2297        & 35.44            & 2.13     & 174 & 12                & 6                  & 95.867338               &    c   \\
HR2343        & 41.73            & 0.92     & 205 & 11              & 23                 & 264.288344              &       \\
HR2356 & 24.98   & 1.5   & 369 & 14 & 4   & 0.002175    &  c, d\\
HR2370 & 23.50   & 1.5   & 379 & 32 & 4   & 0.002465    & c, d\\
HR2490        & 7.22            & 1.92           &  121 & 25    & 9                  & 285.170532              &       \\
HR2532 & 4.25    & 1.5   &   275 &     12 & 3   & 0.001679    & c, d \\
HR2568        & -36.83           & 2.45         &  248   &   16          & 9                  & 469.783901              &       \\
HR2585        & -13.71           & 1.50            &       217   & 15       & 21                 & 960.236042              &       \\
HR26   & -19.44  & 1.5   &    205  &    21 & 6   & 913.119676  & c \\
HR2648        & 27.72            & 2.20            &      338   &    19       & 16                 & 356.01978               &       \\
HR2670 & 22.85   & 1.5       &  261      &   19    & 3   & 0.001759    & c, d \\
HR2763        & -9.86            & 2.75                   &  148      &    6        & 12                 & 156.812257              &       \\
HR2783        & 7.48             & 2.63                  &   286     & 20             & 15                 & 117.72103               &       \\
HR2845        & 17.30            & 3.01                 &    248     &  13            & 50                 & 1739.255972             &  a     \\
HR2944 & 51.47   & 1.5      &    126    &  18       & 4   & 0.002072    &  c, d\\
HR2946 & 9.16    & 1.5        &    150    &  8     & 4   & 788.915614  &  c\\
HR3034 & 35.73   & 1.5       &     412   &   135    & 7   & 303.167615  & c \\
HR3067        & 9.07             & 6.80                      &   158     &    23     & 23                 & 895.712859              &    a   \\
HR311  & -12.38  & 1.5        &       251 &   12    & 7   & 355.911065  &  c\\
HR3131 & -13.81  & 1.5       &      255  &    15    & 3   & 0.001216    &  c, d\\
HR3134        & -10.00           & 1.5                  & 189       &     22        & 6                  & 0.034942                &   c, d    \\
HR3173 & 3.56    & 1.5       &      173  & 24       & 3   & 0.001273    &  c, d\\
HR3192        & 22.01            & 1.09                &    130    &     22          & 14                 & 94.78103                &       \\
HR3474 & 18.04   & 1.5       &     180   &  18       & 6   & 257.361504  & c \\
HR3601        & -17.43           & 1.18                & 160       &     11          & 9                  & 1094.019051             &       \\
HR3662        & -16.13           & 1.30               &    158    &    17            & 10                 & 205.197732              &       \\
HR3665 & -27.14  & 1.5        &    111    &    19   & 3   & 0.001169    &  c, d\\
HR3690        & 4.65             & 1.5                  &    165    &   10          & 6                  & 764.881991              &  c     \\
HR3799        & 23.03            & 4.76                 &   204     &  26            & 35                 & 568.760173              &   a    \\
HR384  & -16.25  & 1.5   &         363  &    33       & 3   & 0.001551    &  c, d\\
HR3858 & 16.05   & 1.5   &       286    &   27         & 3   & 0.00125     &  c, d\\
HR3885        & -5.06            & 1.5          &    320       &   15                  & 6                  & 0.015972                &    c, d   \\
HR3917        & -13.68            & 1.36                   &     144      &    12        & 15                 & 1052.038044             &       \\
HR398  & 8.23  & 1.5    &      124     &   58        & 3   & 1627.157998 &  c\\
HR3982 & 10.81  & 1.5  &        307   &  13         & 21  & 654.254503  &  \\
HR4116        & 18.07            & 0.96        &     130      &  13                     & 25                 & 1143.034294             &       \\
HR4123        & 15.02            & 1.38        &      243     &         17              & 15                 & 754.982338              &       \\
HR4172        & 16.49            & 1.01          &      261     &    12                 & 15                 & 668.142639              &       \\
HR419  & 13.63   & 1.24  &      160     &   23         & 12  & 410.984386  &  \\
HR4259 & -0.49   & 0.98   &       176    &  24         & 13  & 390.999907  &  \\
HR4260 & 1.30    & 0.92   &       219    &    16       & 16  & 521.426806  &  \\
HR4317 & 40.07   & 1.5    &        107   &  34         & 3   & 0.002488    & c, d \\
HR4388 & -5.19   & 1.5   &     255      &       22     & 6   & 91.775648   &  c\\
HR4422        & -9.86            & 2.46              &     181      &   11              & 39                 & 881.752685              &       \\
HR4468        & 2.09             & 3.78              &      204     &  8               & 52                 & 1608.756203             &   a    \\
HR4515        & 0.56             & 1.13               &     138      &     9           & 21                 & 1123.912442             &       \\
HR4787        & -14.89           & 1.27             &      132     &         16         & 12                 & 521.674422              &       \\
HR4828        & 3.23             & 1.61             &      139     &   12               & 36                 & 1181.780093             &       \\
HR4875 & -13.32  & 1.5    &      173     & 27          & 5   & 385.056157  & c \\
HR4886 & -15.14  & 1.5    &        239   &  45         & 5   & 54.920266   &c  \\
HR4936        & -32.45           & 1.5     &   216  &   15              & 7                  & 316.121968              &    c   \\
HR496  & -4.94  & 2.71       &         387    &    58          & 20  & 901.55287   &  \\
HR5037 & -1.87   & 1.5        &      237       &  24            & 7   & 56.930243   &  c\\
HR5062        & -15.01           & 1.02         &      231        &   12                & 13                 & 1100.929109             &       \\
HR5107        & -12.89           & 1.12           &       236       &     15                   & 15                 & 1085.97051              &       \\
HR5112        & -22.90           & 0.83             &      148        &       11               & 30                 & 683.14287               &       \\
HR5127 & -11.58  & 2.12       &         185     & 32           & 4   & 0.001979    & c, d \\
HR5179 & -13.01  & 1.5        &         224     &     21      & 3   & 0.001482    &  c, d \\
HR5238 & -13.13  & 1.5        &       242       &     34      & 4   & 0.002616    &  c, d \\
HR5244        & -23.84           & 1.5                 &    121          &   18               & 7                  & 768.021759              &    c   \\
HR545         & -4.57             & 9.28              &     67         &         11            & 19                 & 864.65926               &    a   \\
HR5478        & 9.05             & 1.36             &         107     &    21                  & 17                 & 776.813275              &       \\
HR5511        & -5.88            & 3.12            &       262       &       14                & 77                 & 2685.67243              &    a   \\
HR5517 & 4.39   & 3.73       &        142      &    10        & 10  & 858.685995  & a \\
HR5685        & -34.76           & 1.95                  &     213         &          12       & 15                 & 19.981123               &       \\
HR5735 & -2.81   & 1.5        &                193   &  57              & 3   & 0.0011      & c, d \\
HR5849        & -15.26           & 4.54                   &         182     & 12               & 6                  & 1574.728785             &   a    \\
HR586         & -27.14           & 2.97                  &          306    &   37              & 9                  & 1970.238947             &       \\
HR5867        & 3.32             & 3.01              &         211     &        8             & 32                 & 1082.189607             &   b    \\
HR5938        & -18.44           & 1.75               &       256       &   49                 & 29                 & 769.839537              &       \\
HR5949 & -9.88  & 1.64       &          162    &  12          & 24  & 764.902535  &  \\
HR6003        & -16.95           & 1.87                &   36           &       22            & 13                 & 29.135775               &       \\
HR6013        & -15.78           & 1.5                &    264          &     16              & 6                  & 0.002523                &    c, d   \\
HR6036        & -10.17           & 1.5              &      166        &      25               & 6                  & 0.002755                &    c, d   \\
HR6051        & -5.77            & 0.99           &     303         &        12                & 12                 & 27.025555               &       \\
HR6054        & -10.66           & 1.5            &      113     &     21            & 6                  & 0.024664                & c, d      \\
HR6110        & 2.65             & 1.5                &    226       &          26                   & 6                  & 24.98375                &    c, d   \\
HR615  & 38.04  & 1.5          &      239     &         28          & 6   & 343.953368  & c  \\
HR6410        & -30.47           & 1.28               &     315      &     22                         & 15                 & 1184.671933             &       \\
HR6502 & -30.78  & 1.5            &        263   &           36      & 3   & 0.001342    & c, d \\
HR6511 & 5.64    & 1.5            &       321    &      23           & 4   & 0.0025      & c, d  \\
HR6534 & -22.46  & 1.5             &      230     &  29              & 3   & 0.001377    & c, d \\
HR6629 & -22.06  & 1.5             &       176    &   16             & 3   & 2645.684549 &  c\\
HR664         & 11.15            & 2.01                         &        237   &   14               & 12                 & 478.720694              &       \\
HR6700 & -13.55  & 2.16             &           155 &          30      & 8   & 57.966077   &  \\
HR6723        & 8.06             & 1.5                            &      168     &             26    & 6                  & 0.981238                &    c, d   \\
HR6747 & 17.23   & 1.5             &     287      &      36          & 6   & 128.712014  & c \\
HR6779        & -30.01           & 1.22                      &     171      &   24                    & 86                 & 2263.793761             &       \\
HR6789 & -7.95  & 1.5             &         196  & 41               & 3   & 0.001412    & c, d \\
HR6826 & -10.09  & 1.5            &      255     &     33            & 5   & 522.625034  &  c\\
HR6827 & -19.56  & 1.03            &     175      & 19                & 36  & 1156.87287  &  \\
HR6873        & -14.94           & 1.5                      &      234     &  17                     & 7                  & 258.202708              &  c     \\
HR6881 & -12.15  & 1.5             &    203       &   36             & 3   & 0.001354    &c, d  \\
HR6923 & -29.82  & 1.5             &        215   &        17        & 6   & 0.026435    & c, d \\
HR6930 & 22.43  & 1.5             &        166   &       37         & 3   & 0.001238    & c, d \\
HR708  & 11.00   & 1.83            &           234 &    21            & 72  & 1385.298357 &  \\
HR7096 & -34.50  & 1.5             &          142  &    17            & 3   & 0.001516    & c, d  \\
HR7142        & -52.88           & 1.5                       &       257    &       44               & 3                  & 0.001216                &    c, d   \\
HR7202 & -23.68  & 1.5             &           264 &  31              & 3   & 0.001319    &  c, d\\
HR7235        & -22.36           & 2.88                      &       280    &        17               & 19                 & 504.788991              &       \\
HR7236        & -7.98            & 0.96                      &    76       &  20                     & 9                  & 539.651655              &       \\
HR7249 & -19.75  & 1.5             &     186      &   42             & 3   & 0.001262    & c, d \\
HR7262 & -24.57  & 1.5             &     259      &      19          & 3   & 0.001354    & c, d \\
HR7403 & -19.39  & 1.5            &       315    &   44              & 3   & 0.001944    &  c, d\\
HR7420        & -22.24           & 2.21                      &       219   &   23                   & 21                 & 1096.953044             &       \\
HR7446        & -20.94           & 1.11                      &       263    &          36             & 16                 & 828.693796              &       \\
HR7457        & -13.98           & 1.92                     &    208       &   79                     & 74                 & 707.014815              &       \\
HR7466        & -16.93           & 2.46                    &        171   &        16                 & 30                 & 706.036806              &       \\
HR7528        & -24.68           & 2.96                     &        168   &        20                & 34                 & 1160.874757             &       \\
HR7543        & -26.43           & 1.47                       &      232     &         21             & 17                 & 133.771922              &       \\
HR7565        & -17.91           & 1.23                       &     209      &      36                & 9                  & 617.270567              &       \\
HR7600 & -56.19  & 1.5           &           278  &           51       & 6   & 377.831123  &  c\\
HR7708        & -1.98           & 11.46                        &        349   &    35                & 64                 & 1060.992003             &   a    \\
HR7724 & -23.48  & 1.5            &       207    &   53             & 6   & 91.049363   & c \\
HR7740 & -16.13  & 2.53           &      222     &       26          & 12  & 828.785996  &  \\
HR7757        & -12.16           & 2.31                        &  184         &   14                  & 18                 & 704.138507              &       \\
HR7803 & -4.82   & 1.5             &      183     &    30            & 4   & 579.418577  & c \\
HR7890 & -10.16  & 1.5             &        246   &      49          & 3   & 0.00162     &c, d  \\
HR7906        & -2.90            & 1.60                       &    124       &           13           & 31                 & 1085.153819             &       \\
HR793  & -3.31   & 1.5   &     162     &   48  & 3   & 0.001308    &  c, d\\
HR7950 & -14.33  & 0.85              &      110     &   7       & 12  & 80.817974   &  \\
HR801         & 9.83            & 5.62                      &       68    &  18                & 27                 & 1119.029144             &   a    \\
HR8028        & -24.61           & 3.35                        &          207 &   7             & 51                 & 1717.179247             &   b    \\
HR804         & -3.84            & 1.5                          &     179      &    30          & 6                  & 389.851169              &   c    \\
HR8047 & 14.42    & 1.5             &       334    &     28      & 4   & 0.002257    &  c, d\\
HR8146        & -1.93            & 0.59               &   230        &      24                   & 23                 & 1107.927408             &       \\
HR8270 & -20.86  & 0.92             & 261          &     29     & 9   & 489.834769  &  \\
HR8319 & 2.35   & 1.5              &         172  & 38         & 3   & 0.001273    &  c, d\\
HR8342        & -1.53            & 0.82                        &        207   &   17             & 12                 & 413.913171              &       \\
HR835  & -11.92  & 1.5          &        180         &    16               & 2   & 0.000729    &  c, d\\
HR8373 & 9.61    & 1.5            &                194  &   40                 & 3   & 0.001343    &  c, d\\
HR838  & -3.01   & 1.21            &           205      &           22        & 202 & 2612.853019 &  \\
HR8402 & 11.50  & 1.5            &              224   &                 25   & 3   & 0.001389    & c, d \\
HR8438 & -49.57 & 1.5            &        201         &             19      & 6   & 955.348137  &c  \\
HR8450 & -10.01 & 0.56            &         159        &         13          & 23  & 152.697998  &  \\
HR8451 & -6.20  & 1.5             &             191    &             35      & 6   & 0.013518    &  c, d\\
HR8597 & -3.69  & 1.92            &             206    &         13          & 9   & 1095.852153 &  \\
HR8628 & 2.01   & 0.75            &           188      &       34            & 6   & 1292.41802  &c  \\
HR8634 & 8.71   & 3.07            &               129  &  8                 & 192 & 2467.214178 & a \\
HR8651        & -14.00           & 1.30     &         142        &       15                & 13                 & 934.67221               &       \\
HR8682        & -12.53           & 1.85     &               295  &         25              & 15                 & 462.639352              &       \\
HR8758        & -18.58           & 1.32     &         318        &            32             & 9                  & 441.771597              &       \\
HR8781 & 0.11  & 1.63          	       &               131  &   9                & 29  & 1541.864433 &  \\
HR879         & 14.88            & 1.35  &                181  &   13                      & 16                 & 707.010613              &       \\
HR8808 & -24.83 & 1.5            &              84   &      42              & 3   & 0.001852    &  c, d\\
HR8936        & 13.89            & 1.74                    &       178          &   10                         & 12                 & 494.629503              &       \\
HR894  & -15.69 & 1.5            &            207     &             53       & 3   & 0.001585    &  c, d\\
HR8976 & -11.87 & 1.53            &             204    &  24                 & 21  & 1289.321494 &  \\
HR8988 & -6.14 & 1.5            &            157     &              46      & 3   & 0.001181    &  c, d\\
HR899  & 20.54  & 1.5             &         127        &           63        & 4   & 81.794676   &  c\\
HR9071        & -9.98           & 1.5                   &   153              &    34                         & 6                  & 901.44463               &     c  \\
HR9098        & 10.65             & 0.96                      &          198       &         21                 & 19                 & 592.253391              &       \\
HR932  & 10.23  & 1.5            &           186      &    43                & 3   & 0.00118     &  c, d\\
HR954  & 21.24  & 1.5             &       78          &           15        & 7   & 17.997014   & c\\
HR980         & 2.58            & 0.73                      &               301  &    38                      & 9                  & 25.973321               &       \\

\enddata
\tablecomments{a, High scatter, discussed in Section \ref{highscatter}. b, Long term trend, discussed in Section \ref{highscatter}. c, Too few measurements, error was artificially assigned to the median value for our method. d, Too short time baseline, error was artificially assigned to the median value for our method. }
\end{deluxetable*}

\end{document}